\documentclass[preprints,article,accept,moreauthors,pdftex]{Definitions/mdpi} 

\usepackage{graphicx}
\usepackage{dcolumn}
\usepackage{bm}
\usepackage[utf8]{inputenc}
\usepackage[english]{babel}
\usepackage{tabu}
\usepackage{float}
\usepackage{braket}
\usepackage{physics}
\usepackage{bbold}
\usepackage{bbm}
\usepackage[title]{appendix}
\usepackage{txfonts}
\usepackage{hyperref}

\usepackage{amsmath}  
\usepackage{latexsym}
\usepackage{amsfonts}   
\usepackage{amssymb}
\usepackage{array}      
\usepackage{epsfig}
\usepackage{txfonts}

\newtheorem{thm}{Condition}

\firstpage{1} 
\pubvolume{xx}
\issuenum{1}
\articlenumber{5}
\pubyear{2019}
\copyrightyear{2019}
\history{}

\Title{Thermalization of finite many-body systems by a collision model}

\Author{Onat Ar{\i}soy,$^{1,2,3,*}$, Steve Campbell,$^{4}$ and \"{O}zg\"{u}r E. M\"{u}stecapl{\i}o\u{g}lu$^1$}

\AuthorNames{Onat Ar{\i}soy, Steve Campbell and \"{O}zg\"{u}r E. M\"{u}stecapl{\i}o\u{g}lu}

\address{$^1$Department of Physics, Ko\c{c} University, \.{I}stanbul, Sar\i yer 34450, Turkey\\
$^2$Institute for Physical Science and Technology, University of Maryland, College Park, Maryland 20742, USA\\
$^3$Chemical Physics Program, University of Maryland, College Park, Maryland 20742, USA\\
$^4$School of Physics, University College Dublin, Belfield Dublin 4, Ireland}

\corres{Correspondence: oarisoy@umd.edu}

\abstract{We construct a collision model description of the thermalization of a finite many-body system by using careful derivation of the corresponding Lindblad-type master equation in the weak coupling regime. Using the example of two level target system, we show that collision model thermalization is crucially dependent on the various relevant system and bath timescales and on ensuring that the environment is composed of ancillae which are resonant with the system transition frequencies. Using this we extend our analysis to show that our collision model can lead to thermalisation for certain classes of many-body systems. We establish that for classically correlated systems our approach is effective, while we also highlight its shortcomings, in particular with regards to reaching entangled thermal states.}

\begin{document}
\section{Introduction}
Computer simulations of finite many body systems have been challenging and expanding predictions of statistical mechanics since their first application to test equilibration of an anharmonic crystal modeled by a chain of masses with fixed-ends~\cite{ford_fermi-pasta-ulam_1992}. While standard methods to investigate equilibration and thermalization of quantum systems are based upon master equations~\cite{book}, so called quantum collision models are introduced as versatile computational tools for simulating and studying open quantum systems~\cite{collint, scarani2002}. The simplest collision model consists of a two-level system undergoing repeated collisions with environment, or ancilla, two-level systems. It is equivalent to a discrete time Markovian master equation in Lindblad form for the dynamics of the system, for short collision times~\cite{collqo}. Here, we address the question of how to generalize the collision models to finite quantum many-body systems for illuminating their thermalization dynamics. 

Intuitively, it is reasonable to obtain a Markovian dynamics of the system using collisions if the colliding ancillae do not interact with any other degrees of freedom since such short time interactions should not allow any significant memory effects. However, the often implicit assumption of stronger interaction than the system Hamiltonian and the neglecting of the bath Hamiltonian are not always valid. Furthermore, using the typical formalism, e.g. Refs.~\cite{collqo, RuariPRA} where energy preserving exchange interactions are considered, results in a dynamics which drives the system to the same state as ancillae, meaning that the result is independent from the system Hamiltonian and homogenization, rather than thermalization, is  achieved~\cite{PezzuttoNJP, BarraSciRep}. This problem persists and is compounded for the generalization of collision models for many-body systems~\cite{compcoll, BarraSciRep}. Interestingly, Ref. \cite{Barra_2017} derives a Lindblad type master equation for collisions with arbitrary interaction strengths and collision times and establishes that the thermal state of a system at the environment temperature with respect to the Hamiltonian $\hat{H}_{0}$ is an equilibrium state if $[\hat{U},\hat{H}_{0}+\hat{H}_{bath}]$ where $\hat{U}$ is the unitary evolution operator under the total Hamiltonian. However, setting $\hat{H}_{0} \!=\! \hat{H}_{system}$ and finding the necessary interaction type and strength to satisfy this commutation property remains as a challenging open problem so far. At variance with this and other works that study collision models starting from a ``global" unitary picture~\cite{compcoll, Barra_2017}, in this work we propose a master equation derivation inspired by the well-known derivation for a time-independent system-bath interaction in the weak coupling regime \cite{book}.

Despite its drawbacks in describing Markovian open system dynamics, quantum collision models are still a good candidate for understanding the quantum thermodynamical phenomena from a microscopic perspective~\cite{DeChiaraNJP}. For example, the microscopic Markovian master equation derivation in Ref.~\cite{book} does not account for the information loss of the system about its initial state, while it is evident using the collision model that the lost information is kept by the entanglement between the system and ancillae~\cite{collinf}. Another study analyzes the entropy generation and distribution in a collision model and proves the asymptotic factorization of the total density matrix of system and environment into the density matrices of the system and the environment for a two level system in the strong coupling regime \cite{Cusumano_2018}. More complex collision models involving ancilla-ancilla collisions allow for the derivation of completely positive non-Markovian dynamics~\cite{nmcoll,nmcoll2, BassanoPRL}. Controllable degree of non-Markovianity and its effect on the dynamics of quantum coherence has been examined~\cite{cakmak_non-markovianity_2017}. Further attempts to study non-Markovian dynamics are made by using initially entangled ancillae~\cite{nmcoll3}, introducing time overlap of two consecutive collisions \cite{RuariPRA, nmcoll4, CampbellPRA2018} and using a two-spin system in which only one of the pair interacts with the environment resulting in a Markovian dynamics for the composite system, while tracing out the spin interacting with the bath gives a non-Markovian dynamics for the remaining spin~\cite{nmcoll5}. The versatility of collision models has resulted in other interesting research directions such as the introduction of collisions with non-thermalized ancillae to study non-equilibrium effects in quantum thermodynamics~\cite{noneq-coll, noneq-coll2, noneq-coll3, noneq-coll4, DeChiaraNJP, BarraSciRep} and the generation of multi-qubit entanglement via a shuttle qubit colliding with disjoint qubit registers~\cite{coll-entg}.

This work aims to examine the conditions required for thermalization in a Markovian collision model using two level ancillae. To this end, we will first carefully examine the microscopic derivation of a Lindblad master equation for a two level system in the weak coupling regime from~\cite{book} and introduce a time dependent interaction Hamiltonian in Sec.~\ref{drv}, where we also assess each assumption made for the derivation and examine their validity. Sec.~\ref{mbs} examines our collision model for many-body systems for both non-entangled and entangled energy eigenstates with an example for each of these cases illustrating how our proposed collisional route to many-body thermalization works. Finally, we conclude in Sec.~\ref{conc}.

\section{Derivation and validity of Lindblad master equation\label{drv}}
We begin by following the microscopic derivation of the Lindblad master equation given in Ref.~\cite{book}, however allowing for a time-dependent interaction Hamiltonian instead of using the second order approximation of the unitary evolution operator for the system and the ancilla with respect to the collision time~\cite{nmcoll5}. The dynamics of the system and the bath is governed by the Liouville-von Neumann equation
\begin{equation}
\frac{d}{dt}\rho(t) = -i[\hat{H}_{I}(t),\rho(t)].
\end{equation} 
Integrating this equation with respect to time and plugging in the expression for $\rho(t)$ in the commutator twice with the assumption of $\text{Tr}_{B}([\hat{H}_{I}(t),\rho(0)])\!=\!0$ we arrive at
\begin{equation}
\frac{d}{dt}\rho(t) = -\int_{0}^{t}ds[\hat{H}_{I}(t),[\hat{H}_{I}(s),\rho(s)]].
\end{equation} 
Applying the Born approximation by neglecting system-bath entanglement and the effect of the system on the bath allows us to write an equation for the dynamics of the system by tracing over the bath degrees of freedom
\begin{equation}
\frac{d}{dt}\rho_{s}(t) = -\int_{0}^{t}ds\text{Tr}_{B}([\hat{H}_{I}(t),[\hat{H}_{I}(s),\rho_{S}(s) \otimes \rho_{B}]]).
\end{equation}
At this point, the dynamics of the system is still, in general, non-Markovian and we have not made any explicit assumptions about the nature of the interaction. However, the finite time of a given collision may serve to justify the constancy of the bath density matrix along with the weak interaction assumption. Putting aside the validity of Born approximation, we need to explicitly assume that the density matrix of the system does not change significantly during the interaction with a single ancilla, which is justifiable for short collision times, in order to replace the past states of the system with its present state and to obtain the Redfield equation
\begin{equation}
\frac{d}{dt}\rho_{s}(t) = -\int_{0}^{t}ds\text{Tr}_{B}([\hat{H}_{I}(t),[\hat{H}_{I}(t-s),\rho_{S}(t) \otimes \rho_{B}]]).\label{me1}
\end{equation}
The standard master equation derivation in Ref.~\cite{book} for a time-independent interaction Hamiltonian continues with the assumption that the integrand above vanishes quickly enough to extend the integral to infinity with negligible difference on the system dynamics. In our case of short time collisions starting after $t = 0$, this extension is not an assumption to be checked as the integrand is explicitly zeroed out for $s\!>\!t$ by the time-dependent strength of the interaction Hamiltonian. For simplicity, we assume that each ancilla interacts with the system once and these collisions start with a period of $\tau_{p}$ and a duration of $\tau_{c}$.

After explicitly defining our collision model, we can investigate the effects of the finite time interactions on the dynamics. As in the derivation in Ref.~\cite{book}, we will introduce the interaction Hamiltonian in the Schr\"{o}dinger picture
\begin{equation}
\hat{H}_{I} = \sum_{\alpha} \hat{A}_{\alpha} \otimes \hat{B}_{\alpha} \label{hi}
\end{equation}
where the Hermitian operators $\hat{A}_{\alpha}$ and $\hat{B}_{\alpha}$ act on the system and the bath respectively. After decomposing the operators $\hat{A}_{\alpha}$ into operators $\hat{A}_{\alpha}(\omega)$ based on the energy transitions with frequency $\omega$ generated on the eigenstates of the system Hamiltonian and plugging the interaction picture interaction Hamiltonian in Eq.~(\ref{me1}), we obtain
\begin{equation}
\begin{aligned}
\frac{d}{dt}\rho_{s}(t)= \sum_{\omega , \omega'} \sum_{\alpha , \beta} & e^{it(\omega'-\omega)} \Gamma_{\alpha\beta}(\omega) \Big( \hat{A}_{\beta}(\omega)\rho_{s}(t) \hat{A}_{\alpha}^{\dagger}(\omega') \\& - \hat{A}_{\alpha}^{\dagger}(\omega')\hat{A}_{\beta}(\omega)\rho_{s}(t) \Big) + \text{h.c.}\label{me2}
\end{aligned}
\end{equation}
where $\Gamma_{\alpha\beta}(\omega)$ is the one-sided Fourier transform of the reservoir correlation functions
\begin{equation}
\Gamma_{\alpha\beta}(\omega) = \int_{0}^{\infty}dse^{is\omega}\text{Tr}_{B}(\hat{B}_{\alpha}^{\dagger}(t)\hat{B}_{\beta}(t-s))
\end{equation}
where operators are defined in the interaction picture. 

For the evaluation of bath correlation spectra, we must specify our open system setup which consists of the same basic ingredients as Refs.~\cite{RuariPRA, nmcoll4, CampbellPRA2018, noneq-coll4}. We first consider a two-level system with time-independent Hamiltonian
\begin{equation}
\hat{H}_{S} = h_{s}\hat{\sigma}_{z}.
\end{equation}
The reservoir consists of, an in principle infinite number of, two-level systems prepared at an inverse temperature $\beta_b\!\!=\!\!1/(k_B T)$ for a bath Hamiltonian
\begin{equation}
\hat{H}_{B} = \sum_{n=1}^{N}h_{b}\hat{\sigma}_{zn}.
\end{equation}
where the index $n$ indicates that the operator acts on the $n^{\text{th}}$ spin of the reservoir. The time-dependent interaction Hamiltonian in the Schr\"{o}dinger picture is given by
\begin{equation}
\hat{H}_{I} = \sum_{n=1}^{N}g_{n}(t)\hat{\sigma}_{x}\otimes\hat{\sigma}_{xn}
\label{interaction}
\end{equation}
where the operator without index acts on the system. For simplicity, we assume that the interaction strength is exactly zero before and after the interaction, remains constant during the collision, and has the same magnitude for all collisions. It should be noted that the interaction in Eq.~\eqref{interaction} is different from the often considered partial swap case which is known to lead to homogenization~\cite{scarani2002} rather than thermalization~\cite{PezzuttoNJP}. Knowing the collision period and duration, we can now define the time-dependent interaction strengths as
\begin{equation}
g_{n}(t) = \theta(t-(n-1)\tau_{p})\theta((n-1)\tau_{p}+\tau_{c}-t)g
\end{equation}
where $\theta$ denotes the Heaviside step function.

Before explicitly calculating the bath correlation spectra, we can make some simplifications. As each ancilla has one interaction component in the form of Eq.~(\ref{hi}), the indices $\alpha$ and $\beta$ in fact denote the index of the corresponding ancilla. Also, knowing that all ancillae are prepared in a thermal state, it is easy to prove that cross correlations vanish and we can arrange Eq.~(\ref{me2}) in the form
\begin{equation}
\frac{d}{dt}\rho_{s}(t) = \sum_{\omega , \omega'} \sum_{n=1}^{N} e^{it(\omega'-\omega)} \Gamma_{n}(\omega) \Big(\hat{A}_{n}(\omega)\rho_{s}(t)\hat{A}_{n}^{\dagger}(\omega') -\hat{A}_{n}^{\dagger}(\omega')\hat{A}_{n}(\omega)\rho_{s}(t)\Big) + \text{h.c.} 
\label{me3}
\end{equation}
After some manipulation, we find the explicit form of reservoir correlation spectra
\begin{eqnarray}
&\Gamma_{n}(\omega) =g^2\theta(t-(n-1)\tau_{p})\theta((n-1)\tau_{p}+\tau_{c}-t)\nonumber\\ &\int_{0}^{\infty}dse^{is\omega}(\rho_{ee}^{n}e^{2ih_{b}s}+\rho_{gg}^{n}e^{-2ih_{b}s})
\theta((t-s)-(n-1)\tau_{p})\theta((n-1)\tau_{p}+\tau_{c}-(t-s))\nonumber\\
&=g^2\theta(t-(n-1)\tau_{p})\theta((n-1)\tau_{p}+\tau_{c}-t)\int_{0}^{t-(n-1)\tau_{p}}ds~e^{is\omega}(\rho_{ee}^{n}e^{2ih_{b}s}+\rho_{gg}^{n}e^{-2ih_{b}s})
\end{eqnarray}
where $\rho_{ee}^{n}$ and $\rho_{gg}^{n}$ are excited and ground populations of $n^{\text{th}}$ ancilla. It is clear that the bath correlation spectra are time-dependent and they are zeroed out by the step functions before or after the collision. We must evaluate this expression for the cases $\omega\!=\! \pm2h_{b}$ and $\omega\! \neq\! \pm2h_{b}$ separately,
\begin{equation}
\Gamma_{n}(\omega,t) = -ig^2(\frac{\rho_{ee}^{n}(\text{exp}(i(t-(n-1)\tau_{p})(\omega+2h_{b}))-1)}{\omega+2h_{b}} + \frac{\rho_{gg}^{n}(\text{exp}(i(t-(n-1)\tau_{p})(\omega-2h_{b}))-1)}{\omega-2h_{b}}),~~~~~\omega \neq \pm2h_{b}.
\label{corr}
\end{equation}
If $\omega\!=\! \pm 2h_b$, one of the complex exponentials in the integrand simplifies and gives a linearly growing term
\begin{equation}
\begin{aligned}
&\Gamma_{n}(\omega,t) =g^2(\rho_{ee}^{n}(t-(n-1)\tau_{p}))-\frac{i\rho_{gg}^{n}(\text{exp}(i(t-(n-1)\tau_{p})(\omega-2h_{b}))-1)}{\omega-2h_{b}},~~~~~\omega = -2h_{b}\\
&\Gamma_{n}(\omega,t) =g^2(\rho_{gg}^{n}(t-(n-1)\tau_{p}))- \frac{i\rho_{ee}^{n}(\text{exp}(i(t-(n-1)\tau_{p})(\omega+2h_{b}))-1)}{\omega+2h_{b}},~~~~~\omega = 2h_{b}
\end{aligned}
\end{equation}

\begin{figure}[t]
	\begin{center}
	\includegraphics[width=0.5\columnwidth]{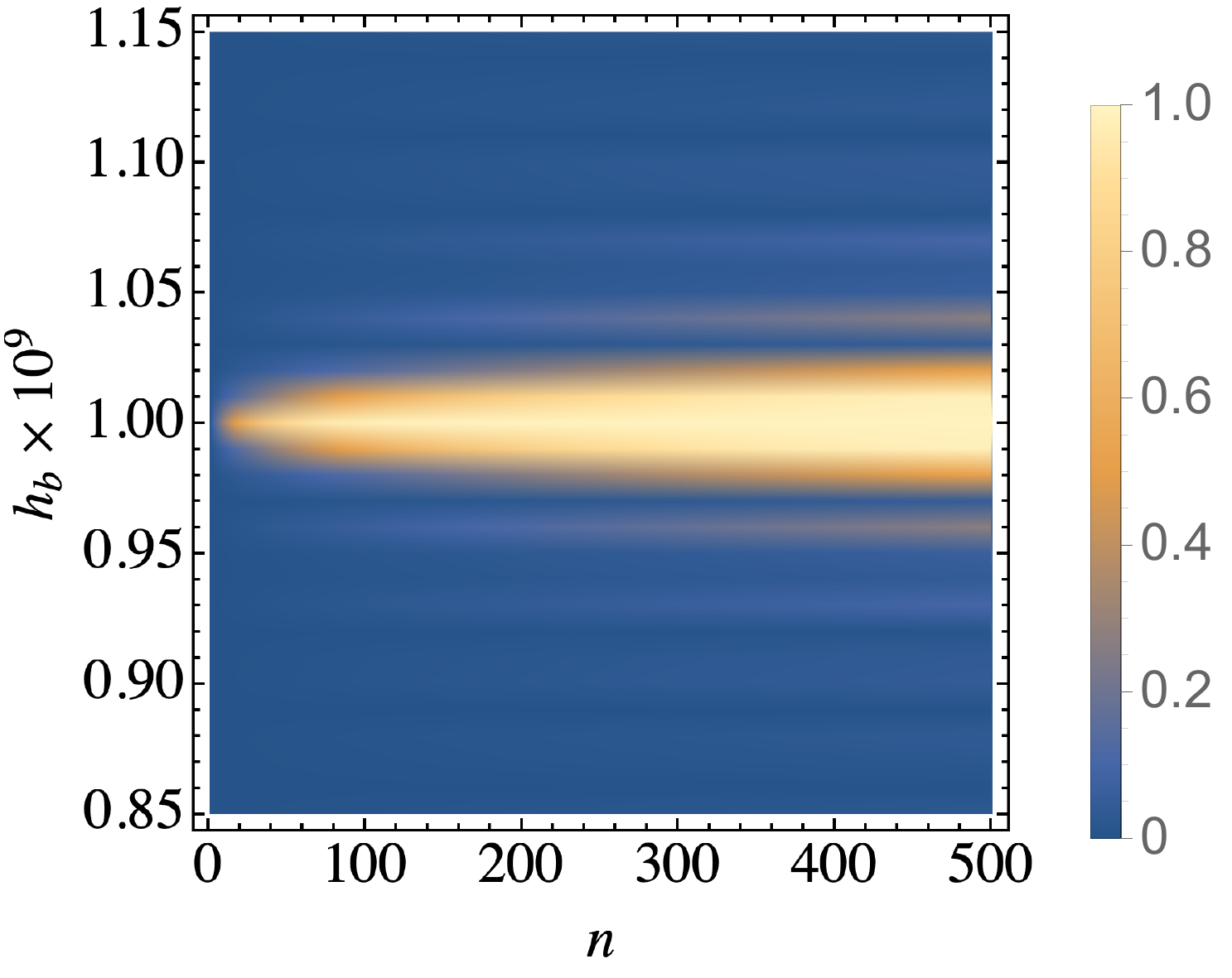}
	\end{center}
	\caption{Simulation results for the thermalization of a single two-level system using our collision model. We show the fidelity of the system with the target thermal state as a function of the bath ancilla frequency and number of collisions. We clearly see that thermalization occurs when the system interacts with bath frequencies that are on resonance. We have fixed $T\!=\!10$ mK, $g\!=\!1$ MHz, $h_s\!=\!1$ GHz, $t\!=\!200$ ns, and $\rho_s(0)\!=\!\ket{1}\!\bra{1}$.}
\label{fig:fidelity}
\end{figure}

The final step of the derivation of a Lindblad type master equation is the decomposition of bath correlation spectra into its real and imaginary parts. The imaginary part results in an additional Hamiltonian term, the Lamb shift acting on the system. However, as this is not relevant to the equilibration of the system, we will neglect it in what follows. As we explicitly show in Fig.~\ref{fig:fidelity} it is also reasonable to neglect situations where ancillae spins are not on resonance with the system, i.e. we only consider $h_b\! = \!h_s$. In this case, the bath correlation spectra consists of a real and linearly growing term and a rotating term with real and complex parts. The linearly growing term generates a dynamics similar to a Lindblad master equation with time-independent interactions, while the real part of the rotating term can be neglected assuming that the relaxation of the system is much slower than the dynamics of the closed system. The master equation in Lindblad form can be obtained after applying these assumptions to Eq.~(\ref{me3}) together with the secular approximation resulting in
\begin{equation}
\frac{d}{dt}\rho_{s}(t) =  \text{Re}(\Gamma(2h_{s},t)) (\hat{\sigma}_{-}\rho_{s}(t)\hat{\sigma}_{+} - \frac{1}{2}\{\hat{\sigma}_{+}\hat{\sigma}_{-},\rho_{s}(t)\}) + \text{Re}(\Gamma(-2h_{s},t))(\hat{\sigma}_{+}\rho_{s}(t)\hat{\sigma}_{-} - \frac{1}{2}\{\hat{\sigma}_{-}\hat{\sigma}_{+},\rho_{s}(t)\})
\label{me-tls}
\end{equation}
where the $\Gamma$ function contains the information about all of the collisions
\begin{equation}
\text{Re}(\Gamma(\omega,t)) = g^2 \sum_{n=1}^{N} (\delta'(\omega-2h_b)\rho_{gg}^{n} + \delta'(\omega+2h_b)\rho_{ee}^{n})(t-(n-1)\tau_{p})
\theta(t-(n-1)\tau_{p})\theta((n-1)\tau_{p}+\tau_{c}-t),
\end{equation}
where $N$ denotes the number of ancilla spins. The function $\delta'(\omega)$ is defined as one for $\omega\!=\! 0$ and zero elsewhere, not to be confused with Dirac delta function. Note that this equation neglects the case where the ancilla is not in resonance with the system and it is used throughout Sec. \ref{mbs}. However, the off-resonance effects in Fig. \ref{fig:fidelity} need to be interpreted using the bath correlation spectrum described in Eq. (\ref{corr}).

The transition from Eq. (\ref{me3}) to Eq. (\ref{me-tls}) takes the secular approximation for granted, however it can be justified by some assumptions relating three different time scales of the open system dynamics: The natural evolution times of the system and ancillae and the duration of the collision, all of which play a critical role in constraining the validity of the derived master equation. We assume that the interaction vanishes before any significant change on the density matrix of the ancilla can happen. We also assume that the variation of the system state during one collision is small, which further constrains the maximum collision time. On the other hand, we also want to eliminate the rotating terms of the bath correlation spectra by averaging them over multiple periods of the system dynamics with a slow relaxation of the system which leads to a lower bound of the collision duration. 

After justifying the derivation of Eq. (\ref{me-tls}), we trivially find the Kubo-Martin-Schwinger (KMS) condition for $n^{\text{th}}$ collision exploiting the fact that the ancillae are prepared in a thermal state, resulting in vanishing cross bath correlations.
\begin{equation}
\frac{\text{Re}(\Gamma_n(2h_{s},t))}{\text{Re}(\Gamma_n(-2h_{s},t))} = \text{exp}(2{\beta}h_{s}) = \frac{\rho_{gg}^{n}}{\rho_{ee}^{n}} = \text{exp}(2{\beta}_{b}h_{s})
\end{equation}
The interpretation of this equation is obvious: The thermal state of the system at the inverse bath temperature $\beta_b$ is the unique steady state of the Markovian dynamics generated by collisions with ancillae prepared in thermal state~\cite{book}. This result was also predicted in complementary works on collision models~\cite{collqo,noneq-coll4} derived using different parameter regimes.


In Fig.~\ref{fig:fidelity} we simulate our collision model sweeping through a range of frequencies for the bath ancillae and show the final state fidelity between the system and its target thermal state. The simulation consists of the unitary evolution of the system and ancillae during the collision time with the sum of system, bath and interaction Hamiltonians described above and the ancillae are traced out after each collision without interacting again with the system or other with ancillae. We clearly see that when the ancillae are close to resonance the collision model leads to thermalization of the system. Conversely, when the ancillae are far detuned from $h_s$ we find the system dynamics are almost frozen. This result can be predicted theoretically by calculating the real part of bath correlation spectrum without assuming resonance. Equation (\ref{corr}) has two terms which are inversely proportional to the difference between the transition frequency $\omega$ and ${\pm}2h_b$. Assuming a small detuning from either $2h_b$ or $-2h_b$, the other term becomes negligibly small. After dropping the small term, evaluating the real part for the other part gives
\begin{equation}
\text{Re}(\Gamma_{n}({\mp}2h_s,t)) = \frac{{\rho}^{n}_{ee,gg}g^2\text{sin}({\delta}t)}{\delta} , \delta = {\mp}2h_s {\pm}2h_b
\end{equation}
ignoring the Heaviside step functions and taking the beginning of each collision as $t = 0$. Its limit for $\delta\rightarrow 0$ recovers the case of resonance. The off-resonance dynamics depend heavily on the product ${\delta}\tau_{c}$. As the average of sine function over a period is zero, we can conclude that the effect of the dissipative term should be negligible if the product ${\delta}\tau_{c}=2k\pi$ where k is an integer and the dynamics is slow enough. On the other hand, in the case where the product is an odd multiple of $\pi$, the average of sine function is not zeroed out and we observe thermalization as seen in Figure \ref{fig:fidelity}. Furthermore, it is trivial to prove that the fastest thermalization is achieved in the case of resonance using the identity
\begin{equation}
\frac{\text{sin}({\delta}t)}{\delta} \geq t, t \geq 0.
\end{equation}

The results in Figure \ref{fig:fidelity} confirm the range of validity of our master equation and are in keeping with other results in the literature~\cite{noneq-coll4}. Furthermore, the clear importance of on-resonance ancillae indicates that, under suitable constraints, only particular bath frequencies are relevant for ensuring the system thermalizes. Thus, we can exploit this feature to explore the requirements for achieving thermalization for many-body systems.

\section{Thermalization of finite many-body systems\label{mbs}}
\subsection{Classically correlated systems}
Let us consider the 1D Ising chain described by the Hamiltonian
\begin{equation}
\hat{H}_S = \sum_{i=1}^{N} h_i\hat{\sigma}_{zi} + \sum_{i=1}^{N-1} J_i\hat{\sigma}_{zi}\hat{\sigma}_{z(i+1)}.
\label{isingH}
\end{equation}
As stressed in the previous section, to achieve thermalization we require the driving frequency of the system and the ancillae to be the same. In the case of interacting many-body systems it should be clear that there will be a range of frequencies, each of which will be related to the various transition frequencies of the many-body system. Thus, to examine the requirements to reach thermalization we use the expression of the interaction Hamiltonian in the form
\begin{equation}
\hat{H}_{I} = \sum_{i=1}^{N}\sum_{n=1}^{N_i}\sum_{\omega}g_{i,n}(t)\hat{\sigma}_{xi}(\omega)\otimes\hat{\sigma}_{x(i,n)}
\end{equation}
where sum over $\omega$ denotes the decomposition of each spin-ancilla collision operator into the different energy transitions it generates. We can make a temporary simplification to make the illustration of many-body system thermalization easier by replacing the ancillae with a set of harmonic oscillators forming a continuous spectrum prepared at an inverse temperature $\beta_b\!=\!1/(k_B T)$. In this case, we can find the energy transitions generated by each term of the interaction Hamiltonian with a partition of the Hilbert space of the whole system based on each nearest neighbor configuration with respect to a reference spin denoted as $i$. We can write all terms of the system Hamiltonian involving $i^{\text{th}}$ spin as
\begin{equation}
\hat{H^i} = (J_{i-1}\hat{\sigma}_{z(i-1)}+h_i+J_{i}\hat{\sigma}_{z(i+1)})\hat{\sigma}_{zi}
\end{equation}
where $i\!\neq\!1,N$ as the first and last spins of the Ising chain do not have a left and right neighbor, respectively. The Hamiltonian at the end points $i\!=\!1, N$ can be found by omitting the term corresponding to the lacking neighbors $i\!=\!0, N+1$ in the above equation.

We can now define the transition frequencies generated by flipping the $i^\text{th}$ spin in terms of the state of neighbor spins
\begin{eqnarray}\label{trf}
\omega\left(\ket{\uparrow^{i-1}\uparrow^{i+1}}\right) &= 2(J_{i-1}+h_i+J_{i}) , \nonumber\\
\omega\left(\ket{\uparrow^{i-1}\downarrow^{i+1}}\right) &= 2(J_{i-1}+h_i-J_{i}) , \nonumber\\
\omega\left(\ket{\downarrow^{i-1}\uparrow^{i+1}}\right) &= 2(-J_{i-1}+h_i+J_{i}) , \\
\omega\left(\ket{\downarrow^{i-1}\downarrow^{i+1}}\right) &= 2(-J_{i-1}+h_i-J_{i}).\nonumber
\end{eqnarray}
Decomposing the operator $\hat{\sigma}_{x}$ as
\begin{equation}
\hat{\sigma}_{xi}=\hat{\sigma}_{-i} + \hat{\sigma}_{+i},
\label{sx}
\end{equation}
we obtain two dissipators for each term of the interaction Hamiltonian. The frequencies in Eq.~(\ref{trf}) correspond to the transitions generated by $\hat{\sigma}_{-i}$ while their negatives correspond to $\hat{\sigma}_{+i}$. Expressing the frequencies as a function of nearest neighbor configuration for each spin results in the master equation
\begin{equation}
\frac{d}{dt}\rho_s=\sum_{i=1}^{N}\sum_{\{s_i\}}\left(\gamma_i(\omega(s_i))D(\rho_s,\hat{\sigma}_{-i}^{s_i})+\gamma_i(-\omega(s_i))D(\rho_s,\hat{\sigma}_{+i}^{s_i})\right).
\label{me-mbs}
\end{equation}
Here, $\{s_i\}$ is a short hand notation for the respective arguments of the frequencies in Eq.~(\ref{trf}), corresponding to the set of basis vectors of the Hilbert space of the neighbor spins of $i^{\text{th}}$ spin. The notation $\hat{\sigma}_{{\pm}i}^{s_i}$ implies that this operator can be decomposed as
\begin{eqnarray}
\hat{\sigma}_{-i}^{s_i} &=& \ket{\downarrow}_{i}\bra{\uparrow}_{i}\otimes\ket{s_i}\bra{s_i}\nonumber\\
\hat{\sigma}_{+i}^{s_i} &=& (\hat{\sigma}_{-i}^{s_i})^{\dagger}
\end{eqnarray}
and $D(\rho,\hat{o})$ is defined by
\begin{equation}
D(\rho,\hat{o})=\hat{o}\rho\hat{o}^\dagger-\frac{1}{2}\{\hat{o}^\dagger\hat{o},\rho\}.
\end{equation}

\begin{figure}[t]
\begin{center}
{\bf (a)}\hskip0.45\columnwidth {\bf (b)}\\
\includegraphics[width=0.48\columnwidth]{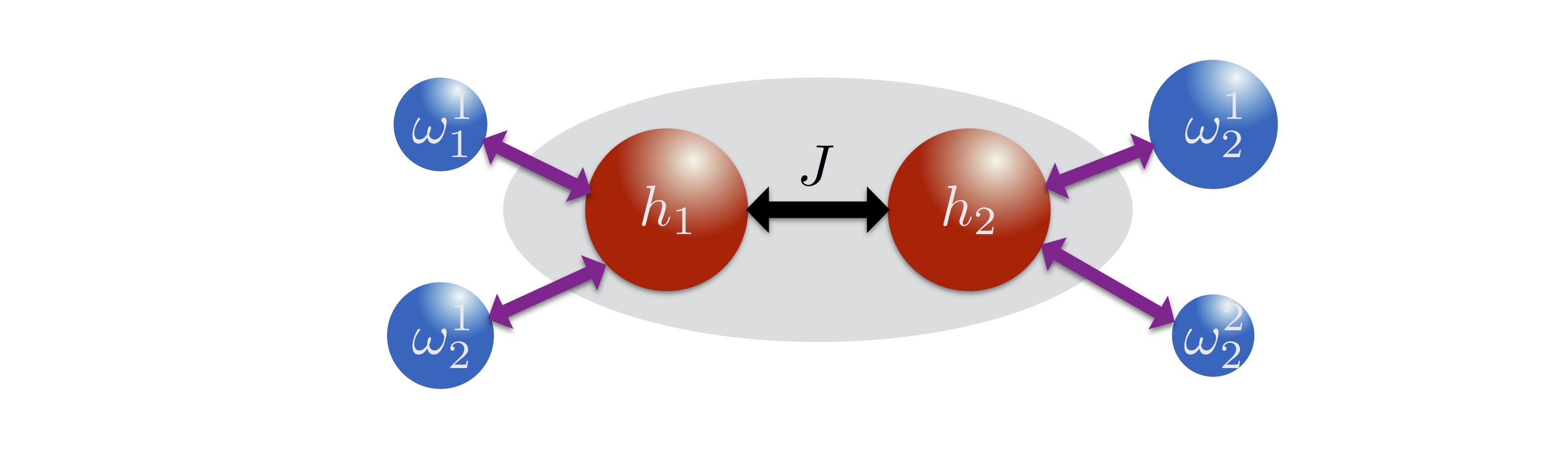}
\includegraphics[width=0.51\columnwidth]{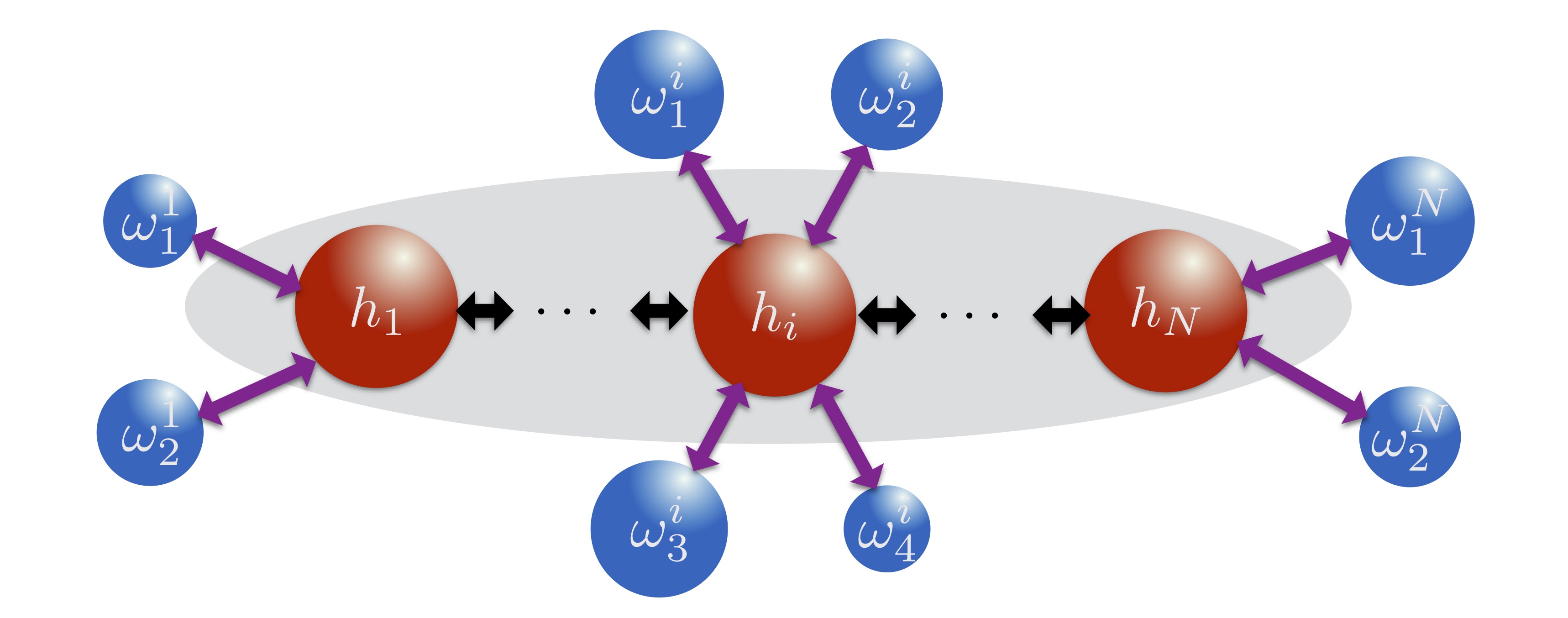}
\end{center}
\caption{Sketch of our proposed collision model thermalizing (a) a two-spin Ising model and (b) a many-body spin model. Complete thermalization requires separate ancillae each corresponding to a spin flip transition frequency of the system.}
\label{fig:coll}
\end{figure}

Although Eq.~(\ref{me-mbs}) is derived for a continuous set of harmonic oscillators, each term appearing in the double sum is similar to the master equation for a two-level system with the driving frequency depending on the nearest neighbor configuration. Therefore, the implementation of a similar master equation with collisions generating one spin flip operations with ancillae driven at the frequencies of single spin transitions, as illustrated in Fig.~\ref{fig:coll}, is possible if the secular approximation is valid such that the ancillae cannot generate any transitions other than those corresponding to its driving frequency. The results of Sec.~\ref{drv} on the KMS conditions for the bath correlation spectra can be generalized for the master equation of 1D Ising model and this ensures that if all ancillae are prepared at an inverse temperature $\beta_b$, the thermal state of the system at the same temperature is a steady state of the master equation~\cite{book}. However, the uniqueness of the stationary solution requires additional constraints. A sufficient condition for the uniqueness can be stated as follows \cite{uniq, thmref}:

\begin{thm}
Let L be the Lindblad superoperator describing the time derivative of the density matrix and $\hat{\sigma}_{{\pm}i}(\omega(s_i))$ operators the generators  of L. The dynamical semigroup generated by L is relaxing in the sense that it drives the density matrix to a unique final state as time tends to infinity regardless of the initial state if the linear span of the generators is an adjoint set and the bicommutant of the generators is the set of all bounded operators acting on the Hilbert space of the system.
\label{thm}
\end{thm}

In order to check the applicability of Condition~\ref{thm} to the thermal bath with local system-bath interactions, we start by checking the adjoint property of the linear span of generators. As established in Ref.~\cite{uniq}, this follows since $\hat{\sigma}_{+i}(\omega(s_i))=\hat{\sigma}_{-i}^{\dagger}(\omega(s_i))$, meaning that the adjoint of each generator is also a generator. The second property is easy to prove using the fact that $\hat{\sigma}_{\pm}$ operators only commute with themselves and the identity operator and the only operator commuting with all $\hat{\sigma}_{{\pm}i}(\omega(s_i))$ for all $i$ and $s_i$ is the identity operator.

To simulate thermalization for a two-site Ising model, Eq.~\eqref{isingH} with $N=2$, we require collisions corresponding to the one-spin flip transition frequencies as illustrated in Fig.~\ref{fig:coll}(a). As each of the spin has a single neighbor, there are two nearest neighbor configurations, resulting in a total of four energy transitions for the whole system. For larger systems, each spin in the bulk of the chain has four different energy transitions and requires more ancillae to successfully thermalize, as shown in Fig.~\ref{fig:coll}(b). 

\begin{figure}[t]
	\begin{center}
	\includegraphics[width=0.5\columnwidth]{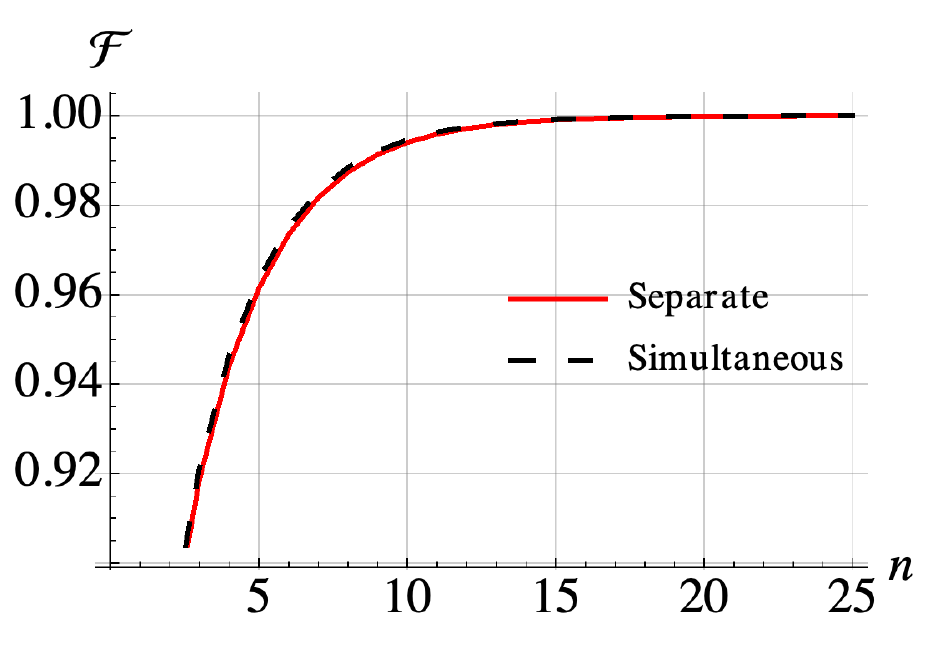}
	\end{center}
	\caption{Simulation of a 2-spin Ising model with parameters $h_1 = h_2 = \text{500 MHz}$ and $J = \text{1 GHz}$ and corresponding transition frequencies $\omega_1=\omega_3=1.5 \text{ GHz}$ and $\omega_2=\omega_4=0.5 \text{ GHz}$. All ancilla-system spin coupling strengths are set as 1 MHz and the collision times are fixed as 400 ns. Fidelity after each step consisting of one collision for each one spin transition frequency with respect to the thermal state of the system at the temperature of ancilla spins $T_{\text{b}} = \text{10 mK}$. The initial state is the thermal state of the system at infinite temperature. }
	\label{fig:2spin}
\end{figure}

We implement our collision model for the two-site Ising chain, considering when the collisions with the various ancillae happen ``sequentially", i.e. the whole system collides with one of the ancillae corresponding to one of the energy transitions at a time and the colliding ancilla is subsequently traced out before the next collision occurs. We also consider ``simultaneously" occurring collisions where the whole system interacts with all of the four ancillae corresponding to different energy transitions at once, after which they are traced out. The minimum energy states are up-down and down-up states and these states cannot be prepared by a local master equation as the collisions are identical, verifying the effect of the system Hamiltonian on open system dynamics resulting in a global master equation. In Fig.~\ref{fig:2spin} we show that our collision model gives rise to thermalization for interacting systems. Furthermore, as the cross bath correlations vanish for a thermal bath, we expect that a time overlap between the collisions (such as that which occurs in the simultaneous collision case) does not change the form of the equation, and our numerical results confirm that both approaches generate an almost identical evolution.

\subsection{Entangled systems\label{apx}}
The Ising model considered in the previous analysis has eigenvectors which are product states without any entanglement among the spin sites. In this section we elaborate on the validity of our collision model for realizing thermalization in more generic many-body systems, particularly those that exhibit entanglement. Addressing such an issue in full generality is a formidable task. Indeed, unlike in the case of non-entangled eigenstates where the generation of single-spin transitions for each interacting neighbor configuration was sufficient, even determining the minimum necessary number of collisions for the uniqueness of the equilibrium state is difficult for entangled states. As such we will restrict to a specific example in this section. 

We begin our discussion by reminding that the matrix representation of any Hamiltonian has an eigenvalue decomposition in the form
\begin{equation}
H_s = UDU^{\dagger}
\end{equation}
where $D$ is a diagonal matrix with the values of eigenenergies on the diagonal, $U$ is a unitary matrix such that its columns are the eigenstates of the Hamiltonian. Following our master equation derivation, each term of the interaction Hamiltonian is decomposed into different energy transitions, giving rise to operators in the form
\begin{equation}
\hat{A}_{kl} = \ket{\psi_{k}'}\bra{\psi_{l}'}
\end{equation}
where $\ket{\psi_{k}'}$ denotes the $k^{th}$ eigenstate of the Hamiltonian, which is also denoted by the $k^{th}$ column of the matrix $U$. This simple form of the energy transition operators can also be expressed in the basis consisting of the Kronecker product of the bases of the subsystems as
\begin{equation}
\hat{A}_{kl} = \sum_{i=1}^{N}\sum_{j=1}^{N} a_{ij,kl}\ket{i}\bra{j}
\end{equation}
where $N$ is the dimension of the Hilbert space of the system and states $i$ and $j$ are selected from the basis constructed by the Kronecker product of the subsystems, therefore these states are not entangled. Knowing that the $k^{th}$ column of the matrix $U$ is equal to $\ket{\psi_{k}'}$, we can write
\begin{equation}
a_{ij,kl} = U^{*}_{ki}U_{lj}
\end{equation}
where $U_{ab}$ denotes the element of $U$ at the $a^{th}$ row and $b^{th}$ column.

The existence of coefficients $a_{ij,kl}$ indicates that there is a one-to-one linear map from the vectors in the basis of eigenstates to the vectors in the Kronecker product basis. Furthermore, we can vectorize the indices $i$ and $j$ into one index $u$ and the indices $k$ and $l$ into another index $v$. By these reductions, we can express our linear map in the form of a matrix $M$ such that
\begin{equation}
M_{uv}A_{v} = x_u
\end{equation}
where the vectors $A$ and $x$ are the vectorized representations of an operator in the basis of eigenstates and in the Kronecker product basis, respectively, with the sum running over the repeated index $v$. As the matrix $M$ represents a one-to-one linear map, its inverse exists and any vector $A_u$ can be expressed in the form
\begin{equation}
M^{-1}_{uv}x_{v} = A_u.
\end{equation}
Using this expression, we can conclude that a single subsystem transition generated by the interaction Hamiltonian, which can be expressed with a vector $x_v$, with one non-zero element can generate multiple energy transitions by the multiplication by the inverse of the matrix $M$ used for conversion into eigenstate basis. In this case, the thermalization conditions depend on the structure of the matrix $M$, however the thermalization of any many-body system is in principle possible with a sufficient number of energy transitions generated by the collisions with two-level ancillae driven at the corresponding transition frequencies, the appropriate choice of interaction Hamiltonian, and the validity of our assumptions for the master equation.

\begin{figure}[t]
	\begin{center}
	\includegraphics[width=0.5\columnwidth]{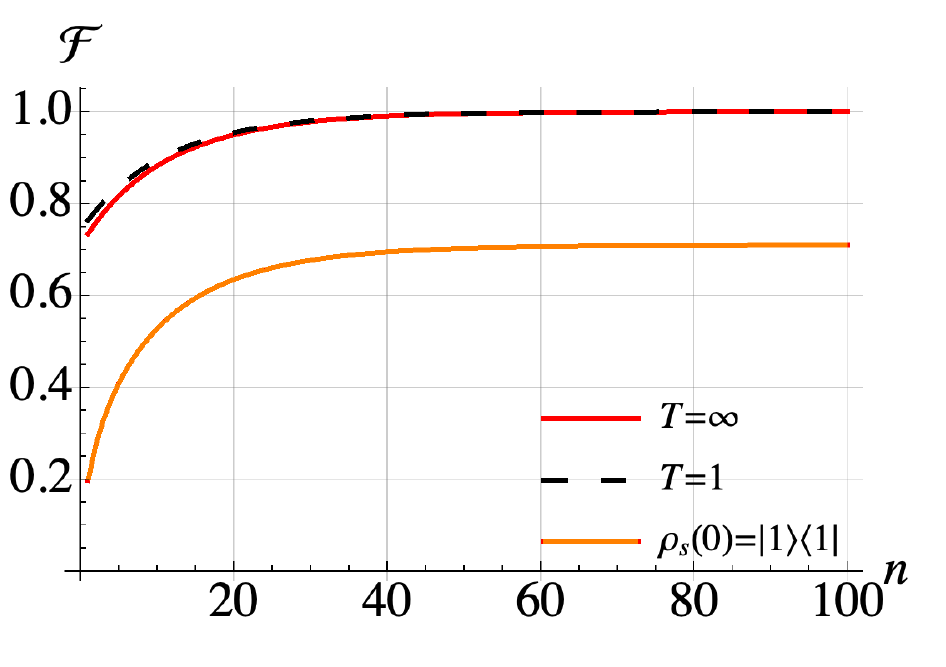}
	\end{center}
	\caption{Simulation of a 2-spin anisotropic $XY$ model with $J = \text{1 GHz}$ with different initial states of the system. Collision time is set as 400 ns with an ancilla-system interaction strength of 1 MHz. Fidelity after each collision between an ancilla driven with the sole non-zero transition frequency of the system $\omega=\text{4 GHz}$ and the first spin of the system each one spin transition frequency with respect to the thermal state of the system at the temperature of ancilla spins $T_{\text{b}} = \text{10 mK}$.}
	\label{fig:xy}
\end{figure}

As a concrete example consider a two-spin anisotropic $XY$-model with Dzyaloshinskii-Moriya (DM) interaction in $z$-direction
\begin{equation}
\hat{H}_{XY} = J(\hat{\sigma}_{x1}\hat{\sigma}_{x2} - \hat{\sigma}_{y1}\hat{\sigma}_{y2} + \hat{\sigma}_{x1}\hat{\sigma}_{y2} - \hat{\sigma}_{y1}\hat{\sigma}_{x2})
\end{equation}
with the eigenstates and eigenenergies~\cite{xx}
\begin{equation}
\begin{aligned}
&\ket{\psi_{1,2}} = \frac{\ket{\downarrow\downarrow} \pm \ket{\uparrow\uparrow}}{\sqrt{2}},~~
&\ket{\psi_{3,4}}= \frac{\ket{\downarrow\uparrow} \mp i \ket{\uparrow\downarrow}}{\sqrt{2}};\\
&E_{1,3} = 2J,~~~~~E_{2,4} = -2J.
\end{aligned}
\label{xxeig}
\end{equation}
Using the definition of operators $\hat{A}_{kl}$, we can express the one spin flip operators as
\begin{eqnarray}
&\ket{\downarrow\downarrow}\!\bra{\uparrow\downarrow} &= -i(\hat{A}_{13} + \hat{A}_{23} - \hat{A}_{14} - \hat{A}_{24})/2\nonumber\\
&\ket{\uparrow\uparrow}\!\bra{\uparrow\downarrow} &= i(\hat{A}_{13} - \hat{A}_{23} + \hat{A}_{14} - \hat{A}_{24})/2\nonumber\\
&\ket{\downarrow\downarrow}\!\bra{\downarrow\uparrow} &= (\hat{A}_{13} + \hat{A}_{23} + \hat{A}_{14} + \hat{A}_{24})/2\nonumber\\
&\ket{\uparrow\uparrow}\!\bra{\downarrow\uparrow} &= (\hat{A}_{13} - \hat{A}_{23} + \hat{A}_{14} - \hat{A}_{24})/2
\end{eqnarray}
We can then describe the spin ladder operators acting on the first site as
\begin{eqnarray}
&\hat{\sigma}_{-1} &= \ket{\downarrow\uparrow}\bra{\uparrow\uparrow} + \ket{\downarrow\downarrow}\bra{\uparrow\downarrow} \nonumber\\
&&=\frac{1}{2}(\hat{A}_{31} - \hat{A}_{32} + \hat{A}_{41} - \hat{A}_{42} +i( -\hat{A}_{13} - \hat{A}_{23} + \hat{A}_{14} + \hat{A}_{24}))\nonumber\\
&\hat{\sigma}_{+1} &= \hat{\sigma}^{\dagger}_{-1}.
\end{eqnarray}
It is clear that the $\hat{A}_{41}$ and $\hat{A}_{32}$ terms of the ladder operators and their Hermitians generate state transitions with non-zero energy difference. Other state transitions are between the states with the {\it same} energy which cannot be generated via with collisions with ancilla spins which do not have internal energy as we have assumed $h_s,h_b\!>\!\!>\!g$. If the zero energy transitions were allowed, we could make transitions from any state of the system to another state using intermediate transitions, impling the uniqueness of the thermal state as the equilibrium point of the dynamics~\cite{ggl}. In our case, this condition is not satisfied, and this leads to the equilibrium state of the system exhibiting an initial state dependence. Our numerical simulations in Fig.~\ref{fig:xy} show that if the system is initially prepared in some thermal state, but not in equilibrium with the bath, a Gibbsian thermal state at the environment temperature is achieved. However, it is not guaranteed for generic non-equilibrium initial states, such as $\ket{1}\!\bra{1}$. We understand this as follows: the choice of initial state as a thermal state at some temperature guarantees that the population of the states having the same energy is equal, thus implying that the zero frequency transition terms will not contribute to the dynamics of the system even if they are generated by the collisions. This means we can assume that the zero frequency terms exist and consequently the equilibrium state is the thermal state at the environment temperature.

Another possible issue regarding thermalization of entangled many-body systems by our collision model is the additional terms of the master equation due to the non-vanishing bath cross correlations arising due to the decomposition of each term of the interaction Hamiltonian acting on a single subsystem into multiple energy transition terms, which implies that the bath operator of those energy transition terms are the same. For this reason, the positive definiteness of the bath correlation matrix for every frequency needs to be asserted for the uniqueness of the equilibrium state \cite{uniq}. 

In summary, our example of two spin anisotropic $XY$ model shows that our collision model can generate multiple energy transitions without the explicit calculation of the $M$ matrix. Although thermalization is not guaranteed, this analysis nevertheless provides insight about how an entangled many-body system with non-degenerate energy levels can be thermalized as long as the secular approximation used in the master equation derivation remains valid and the bath correlation matrix is positive definite.

\section{Conclusion}
\label{conc}
In this work we have presented a collision model using two level ancillae that leads to thermalization in the weak coupling regime, even for certain finite many-body systems. By carefully assessing the relevant timescales present, we showed that when the ancillae are tuned inline with the transition frequencies of the system, thermalization can be achieved. This is at variance with other schemes commonly examined in the literature where system and environment interact via a partial swap~\cite{RuariPRA, CampbellPRA2018}. Our master equation derivation for 1D Ising model can be straightforwardly generalized to $N$-dimensional spin lattices by redefining the sums over the Hilbert space of neighbor spins. In the case of Ising spin lattices with more than one dimension, the system can be tuned to be an integrable or non-integrable system depending whether the external magnetic fields are turned off or on respectively~\cite{integ} and our collision model achieves thermalization in both of the cases. If the eigenstates of the system Hamiltonian are entangled, our collision model gives valuable insight on the dependence of equilibrium state on the initial condition; in particular reveals the conditions to engineer Gibbsian thermal state at the environment temperature. Remarkably, for entangled eigenstates, the decomposition of single-spin transition operators into multiple energy transition operators may remove the necessity of bath interaction with each spin in the system.

Beyond the clear interest in understanding the phenomenology of thermalization using a collision model and its possible extensions to non-Markovian and non-equilibrium dynamics, our collision model also can be viewed as a versatile and implementable artificial environment acting as a temperature knob, as similarly considered in Ref.~\cite{ggl, arXiv2019}. Contrary to the artificial temperature knob proposal in Ref. \cite{ggl}, our proposal satisfies the KMS condition for thermalization instead of an optimized approximation depending on tunable system parameters and it is promising to be scalable for large many body systems. The proposal in Ref. \cite{arXiv2019} relies on a similar idea to our proposal; its authors propose to sweep all possible energy transitions of the system with a slowly varying bath Hamiltonian strength, which can be considered as a different way of obtaining the effect of ancillae colliding to a subsystem with different bath Hamiltonian strength. Obviously, making use of only relevant transition frequencies leads to much faster thermalization and it is possible to get rid of some timescale constraints of Ref. \cite{arXiv2019} as the ancillae are supposed to be prepared in a thermal state for a time independent bath Hamiltonian before the collision in our proposal.

Our proposal can also lead to the cooling of the target system if it is possible to keep ancilla spins colder than the environment temperature. Indeed we mention two possible methods of spin cooling for the preparation of a cold environment that our scheme is well suited to. The first one is the use of frequent measurements on a two-level system interacting with a non-Markovian environment which brings the mean energy of interaction Hamiltonian to zero in order to reduce the total energy of the two-level system and its environment~\cite{meas-cool}. The application of this idea may suffer from the challenges posed by the necessary minimum frequency of the measurements. Another idea is to use quantum coherent or entangled two-level systems~\cite{cakmak_thermal_2017,ent-dim,ozaydin_work_2018} to engineer the  temperature of a two-level target system, which can then be used as an ancilla for the many-body system to be thermalized.

Our results can have practical significance for suggesting design principles of quantum thermalizing machines for finite many-body systems. Such devices would be compact as they can consist of few ancillae as artificial environment; they would be fast as they can engineer the target thermal state with high fidelity after a small number of collisions describing a unitary route to thermalization. These properties can be valuable for quantum thermal annealing~\cite{ggl} and quantum simulation applications~\cite{mostame_quantum_2012}, for example using superconducting circuits.

\authorcontributions{ \"{O}.E.M. and O.A. are responsible for the conceptualization, the formulation of methodology and the theoretical analysis of the ideas introduced in the paper. S.C, \"{O}.E.M. and O.A. equally contributed to the discussions throughout the work, the numerical simulations and the writing of the paper.}

\funding{S. C. gratefully acknowledges the Science Foundation Ireland Starting Investigator Research Grant ``SpeedDemon" (No. 18/SIRG/5508) for financial support.}

\acknowledgments{The authors thank Mauro Paternostro and Giacomo Guarnieri for useful discussions and feedback for early drafts of the paper.}

\conflictsofinterest{The authors declare no conflict of interest.}

\reftitle{References}

\bibliography{collision_turkey}
\end{document}